# Molecular dynamics of a κB DNA element: Base flipping *via* cross-strand intercalative stacking in a microsecond-scale simulation

Cameron Mura[1,*,†] & J. Andrew McCammon[1,2]

## Author affiliations & correspondence:

1 Department of Chemistry & Biochemistry and Center for Theoretical Biological Physics; University of California, San Diego; La Jolla, CA 92093-0365

2 Howard Hughes Medical Institute and Department of Pharmacology; University of California, San Diego; La Jolla, CA 92093-0636

* To whom correspondence should be addressed:
Tel: 1.434.924.7824
Fax: 1.434.924.3710
Email: cmura@virginia.edu

† Current address:
Department of Chemistry; University of Virginia; Charlottesville, VA 22904-4319

## Manuscript information:

| | |
|---|---|
| Synopsis: | Microsecond-scale molecular dynamics simulations of DNA reveal unanticipated conformational features, with implications for site-specific binding of NF-κB transcription factors to regulatory κB DNA elements, as well as spontaneous base flipping *via* a mechanism of cross-strand intercalative stacking of nucleotides. |
| Format: | *Nucleic Acids Research* article (*in press*) |
| Running title: | *Molecular dynamics of κB DNA* |
| Last modified: | 02 July 2008 |
| Length estimate: | 12.5 printed pages (using '*words*/925 + *figures*/2.8'; excluding references) |
| Additional notes: | This manuscript is accompanied by nine figures and thirteen items of supplementary information (nine figures and four video animations). |

## Abstract

The sequence-dependent structural variability and conformational dynamics of DNA play pivotal roles in many biological milieus, such as in the site-specific binding of transcription factors to target regulatory elements. To better understand DNA structure, function, and dynamics in general, and protein···DNA recognition in the "κB" family of genetic regulatory elements in particular, we performed molecular dynamics simulations of a 20-base pair DNA encompassing a cognate κB site recognized by the proto-oncogenic "c-Rel" subfamily of NF-κB transcription factors. Simulations of the κB DNA in explicit water were extended to microsecond duration, providing a broad, atomically-detailed glimpse into the structural and dynamical behavior of double helical DNA over many timescales. Of particular note, novel (and structurally plausible) conformations of DNA developed only at the long times sampled in this simulation – including a peculiar state arising at ≈ 0.7 μs and characterized by cross-strand intercalative stacking of nucleotides within a longitudinally-sheared base pair, followed (at ≈ 1 μs) by spontaneous base flipping of a neighboring thymine within the A-rich duplex. Results and predictions from the μs-scale simulation include implications for a dynamical NF-κB recognition motif, and are amenable to testing and further exploration *via* specific experimental approaches that are suggested herein.

## Introduction

DNA is often viewed as a relatively rigid biological macromolecule (1-3), with RNA and proteins thought of as exhibiting broader ranges of both intrinsic three-dimensional structural variability as well as dynamical flexibility. This perspective of a locally-rigid, globally-flexible biopolymer is consistent with the rather passive biological role of DNA as the repository of genetic information – the genome is *read-out* by the process of transcription. Links between structure and potential biological functions (both normal and aberrant) have been explored for conformations that deviate from the standard *B*-form double helix – including such varieties as multi-stranded triplexes, quadruplex structures found in telomeric G-rich tracts, cruciforms adopted by inverted repeats, hairpins and slipped structures, and so on (4). However, beyond these alternative secondary structures, it is also becoming increasingly apparent that the structure and dynamics of the canonical Watson-Crick DNA double helix on a very local (base pair) level play pivotal roles in specific biological functions, such as the site-specific binding of transcription factors to target DNA elements. This idea of a functional role for sequence-specific DNA fine structure and dynamics is embodied in the concept of "indirect readout," (5,6) wherein features of protein···DNA recognition are dictated by subtle conformational and dynamical properties of DNA beyond the stereochemical code provided by the specific linear array of chemical functionalities that line the major and minor grooves for a given nucleotide sequence.

Despite the vast literature dedicated to DNA structural biology since the first atomic-resolution crystal structures of both left- (7) and right-handed (8) double helices, many aspects of DNA structure and dynamics remain unclear – including the intrinsic coupling between structure and conformational dynamics that is the basis of indirect readout. For instance, controversy surrounds the relative significance of extrinsic/environmental factors (such as hydration and counterion-binding) versus intrinsic factors (such as local base pair interactions) in mediating sequence-specific DNA fine structure, as gauged by helical axis bending, groove widths, and related properties (9-12). Quantitative frameworks have been developed for the description of DNA structure in terms of the local geometry of bases, base pairs (bp), bp steps, and higher-order structural units (*e.g.*, (13)), but progress in elucidating those structural and dynamical phenomena thought to occur *via* transient, short-lived intermediates (such as occurs in base flipping (14,15)) remains hindered by the difficulty of using existing experimental methods to extract dynamical information at both atomic resolution and over the potentially relevant timescales (ns→ms). Emblematic of this difficulty, crystallographic models generally represent spatially- and temporally-averaged structures of individual molecules, the averages being taken over more than $10^{12}$ unit cells (a conservative estimate, for μm-sized crystals of typical cell dimensions) and time periods greater than hundreds of milliseconds (a conservative estimate, for exposure with high-brilliance synchrotron x-rays). Thus, detailed knowledge remains somewhat obscure of the factors modulating the mean structural and conformational properties of DNA, the dynamical processes mediating inter-conversions between these average conformational states, and the interactions of these (sub-)states with transcription factors, histone proteins, intercalators, groove-binding drugs, and other relevant species. Computational approaches such as molecular dynamics (MD) simulation (16) provide an alternative route towards exploring biomolecular structure and dynamics in fully atomic detail, and have yielded a wealth of nucleic acid simulations over the past dozen years (reviewed in (17)).

The NF-κB transcription factor family illustrates the many potential complexities of protein···DNA recognition. This family occurs in a wide variety of eukaryotes and regulates a similarly broad array of cellular pathways, ranging from morphogenesis in insects to adaptive immunity in humans (18). The five mammalian NF-κBs (p50, p52, p65 (RelA), c-Rel, and RelB) contain an ≈ 300-residue Rel Homology Region (RHR), consisting of two immunoglobulin-like folds joined by a flexible linker. NF-κBs associate into homo- and hetero-dimers that modulate gene expression by binding to target κB DNA enhancer sites. The remarkably loose κB consensus sequence $^{5'}$GGGRNWYYCC$^{3'}$ (N = any nucleotide, R = purine, Y = pyrimidine (often Thy), and W = Ade or Thy) consists of two "half sites" (underlined). The 5-bp GGGRN half sites are preferentially bound by p50 and p52 subunits, while RelA, RelB, and c-Rel prefer the 4-bp YYCC half sites. Thus, in addition to immense sequence variability and intrinsically different NF-κB–binding preferences for different half sites, κB elements also vary in length. Known κB sites can be grouped into 9-bp "class I" sites (4+1+4 arrangement, for c-Rel and RelA homodimers) and 10- or 11-bp "class II" sites (5+1+5, for p50 and p52 homodimers). However, even the above rules and consensus sequences are likely to be too restrictive: Some sequence-specific trends are known, but there is great degeneracy in terms of both *(i)* optimal DNA sequences for a given NF-κB dimer and *(ii)* the relative affinities of different NF-κB dimers for a given κB DNA sequence (18). Thus, an outstanding question in the NF-κB field is the detailed mechanism of indirect readout – What are the determinants of sequence-specific binding of NF-κB to target κB sites? Biophysical studies have demonstrated that NF-κB···DNA binding affinity is largely entropically-driven (19), but the issue of site specificity remains far murkier, with it now thought that "the conformation and flexibility of κB DNA sequences play a critical role in the recognition of NF-κB dimers" (20).

Therefore, as an initial step in elucidating protein···DNA recognition and the mechanism of indirect readout in the context of differential binding of NF-κB transcription factors to κB DNA elements, we performed MD simulations of a 20-base pair κB DNA of known structure (Fig. 1, S1). This DNA duplex consists of the sequence d(GGGTTTAAAGAAATTCCAGA), and encompasses a κB element (underlined) recognized by the c-Rel NF-κB homodimer and its oncogenic variant "v-Rel" (21). Simulations of the DNA were extended to the μs timescale, affording insights that would have remained undiscovered in a shorter simulation. Unanticipated κB DNA structural transitions discovered at the long times sampled in this trajectory include cross-strand intercalative stacking ("XSIS") of nucleotides followed by spontaneous base flipping at a neighboring nucleotide, as well as a peculiar minor groove-bound "barbed" terminus. In addition to illuminating the μs-scale structural and dynamical behavior of this particular κB sequence, the simulation provides a broad, atomic-resolution glimpse into the dynamical properties of two turns of double helical DNA over a wide range of timescales (nine orders of magnitude). The terascale body of data presents opportunities for detailed analyses of methodological issues (such as the approximations inherent in the empirical force fields used in MD), but the present report focuses instead upon the intriguing XSIS, flipping, and barbing transitions, as well as suggesting specific experimental ideas which could be used to test and further explore the simulation-based predictions. While technical issues pertaining to force field parameterization lie beyond the scope of the present work, an initial assessment of the quality of this extended trajectory was made by considering μs-scale DNA backbone dynamics in terms of potential $B_I/B_{II}$ and $\alpha/\gamma$ backbone substate sampling problems.

## Materials and Methods

The work proceeded in stages of *(i)* system selection, construction, and preparation; *(ii)* calculation of classical MD trajectories; and *(iii)* data processing and analysis. These stages are described below, and further detailed in the *Supp. Info.* The starting structure was a 20-bp DNA encompassing the κB site of a CD28 response element, drawn from a crystal structure of this duplex bound to the $(\text{c-Rel})_2$ NF-κB homodimer (21). This 9-bp class II κB DNA consists of 4-bp "half sites" (Fig. 1a, S1). Because the dynamics of this particular DNA are of interest as part of a broader NF-κB–related project, the structure of c-Rel–bound κB DNA was used directly and not rebuilt into canonical *B*-form. A virtue of the 20-bp duplex used in these studies (versus a shorter fragment limited to just the κB site) is that the region of primary interest – the 9-bp κB site – is embedded in two helical turns of DNA, making it less susceptible to potentially spurious end-effects. The simulation system was constructed by immersing the duplex in a truncated octahedron of explicit water (Fig. 1b, S1). As described in the *Supp. Info.*, design of the periodic boundary conditions accounted for the calculated hydrodynamic properties of this DNA fragment (Fig. S1e), such that a minimal clearance of 20 Å was maintained between periodic images at all times (this distance being 35 Å for the starting configuration). The final system was prepared *via (i)* addition of hydrogens; *(ii)* successive rounds of energy minimization of protons and DNA atoms; *(iii)* addition of 38 $Na^+$ counterions; *(iv)* placement of an initial shell of interstitial waters around the neutralized DNA•$Na^+$ system; and *(v)* addition of bulk water and ions ($Na^+$ and $Cl^-$). The resultant (over-sized) rectangular cell was trimmed to yield the final mecon. The final 61,439-atom system consisted of the κB DNA duplex accompanied by *(i)* 20,030 molecules of TIP3P water (22); *(ii)* 38 $Na^+$ counterions; and *(iii)* an additional 20 $Na^+$ and 20 $Cl^-$ ions, providing a final [NaCl] ≈ 50 mM (chosen so as to mimic the conditions of DNA-binding experiments with this and related κB elements (23)). As a final preparatory step, the DNA was allowed to adapt to the aqueous ionic environment (and, likewise, the solvent structure to relax around the DNA and remove unfavorable contacts) by a multi-stage protocol involving incremental relaxation of harmonic restraints on DNA atoms over a total of 10,000 cycles of potential energy minimization.

Standard methods were used to compute MD trajectories (see, *e.g.*, (24)). Initial equilibration stages of heating and restrained dynamics (Fig. S2 and the submatrix in Fig. 2a) were followed by 1,020 ns of unrestrained, production-level dynamics. Simulations were performed in the isobaric-isothermal ensemble (constant NPT). Pressure and temperature were maintained at 1 atm and 300 K, with temperature controlled *via* Langevin dynamics for all non-hydrogen atoms, and pressure regulated *via* a hybrid Nosé-Hoover Langevin piston (25). Long-range electrostatics were treated by the smooth particle mesh Ewald algorithm (26), with a grid density better than 1/Å in every direction. Non-bonded short-range interactions were calculated within a spherical cutoff of 10.0 Å (a smooth switching function was applied from 9→10 Å). The SHAKE algorithm (27) was used to constrain bonds between hydrogens and parent heavy atoms, enabling a 2.0 fs integration time step without compromising either bulk thermodynamic quantities (Fig. S2) or structural stability of the trajectory (Fig. S3). Further computational efficiency was afforded using a well-established multiple time-stepping scheme (28). Simulations were computed in 1.0-ns bins, with trajectory coordinates written every 1.0 ps. The trajectory was extended to the μs timescale *via* scripts for automatically linking time slices $i \rightarrow i+1$, enabling indefinite propagation of the simulation in a semi-autonomous manner on a host of available Linux clusters. Trajectories were computed using NAMD (29) and the parm94 force field. Of the many force fields commonly used in biomolecular simulations, the parm94 version of the standard Amber force field was chosen for the sake of consistency and comparability with the wealth of existing DNA simulations that used this particular parameter set. Issues of force field selection and usage present a "moving target," due to continuous developments in this field, and it may be argued that extensive conformational sampling (such as is achieved on the μs timescale) overcomes potential artifacts that may arise in shorter simulations – *i.e.*, sampling and artifact are somewhat orthogonal and counterbalanced effects, and what may appear as "artifact" at short (ns) times is effectively "washed-out" in the limit of longer (μs) timescales. Further details pertaining to force field selection, analysis steps, and issues specific to a simulation of this length are provided in the *Supp. Info.*

## Results and Discussion

The results are presented in five main areas: μs-scale trajectory stability, the cross-strand intercalative stacking transition, a base flipping event, a "barbed" terminus, and trajectory validation in terms of μs-scale sampling of DNA backbone substates. As with any simulation-based work, our computational findings could be viewed as more suggestive than conclusive, and can be considered as a way to generate atomically-detailed hypotheses about the dynamical mechanism underlying a complex biomolecular transition (in this case, XSIS and flipping). Thus, explicit connections to experiment are drawn in two distinct ways (primarily towards the end of each subsection): *(i)* Existing data which bear upon particular results are interleaved into the discussions of those results, as are *(ii)* Specific proposals for experiments that might be suitable in testing our predictions.

### Trajectory stability and conformational variability on the μs timescale

Simulations of the κB DNA element (Fig. 1, S1) were extended past one μs. This lengthy timescale was achieved using an optimized code exhibiting particularly efficient parallel scaling (NAMD; ref. (29)), in conjunction with optimal hardware/network architectures and scripts designed to propagate simulations in a minimally-supervised and essentially uninterruptible manner. Trajectories were thereby calculated over extensive periods of wallclock time (nearly one year), yielding roughly one terabyte of DNA simulation data for this >60,000-atom system. Persistence of the structural integrity of the 20-bp duplex on the μs timescale is notable (Fig. S3), as MD simulations are typically performed for DNA fragments shorter than the two helical turns reported here (*e.g.*, dodecamers), and are generally limited to shorter durations (*e.g.*, <20-30 ns); exceptions are recent studies of counterion-binding in a 60-ns simulation by Beveridge and co-workers (12), two 50-ns trajectories of a different dodecamer by Várnai & Zakrzewska (30), and the recent μs-scale Dickerson dodecamer trjactories used in force field parameterization studies (31). In terms of simulation reliability and ion-related force field artifacts, it should be noted that the recently-studied problem of KCl aggregation as a result of possibly imbalanced Lennard-Jones parameters (32) was not found to be an issue in the simulations reported here; this is not an unexpected result, given the comparatively low (≈ 50 mM) NaCl concentration in this κB DNA system.

The many approximations and assumptions which underlie empirical force fields for biomolecular simulations – simplified potential energy functional forms, parameterizations against short simulations, neglect of atomic polarizability, *etc.*) – motivated careful monitoring of this unusually long κB simulation, in terms of both thermodynamic properties (Fig. S2) and structural parameters such as coordinate root-mean-square deviations (RMSD; Fig. S3). Successful equilibration and stabilization is evident over both short (ns) and long (μs) times. Although RMSD is a less than ideal metric of DNA structural similarity, comparisons of each trajectory snapshot to canonical *A*- and *B*-form DNA suggest the presence of relatively long-lived (>5-ns) states that are closer in structure to *A*- than *B*-DNA (Fig. S3f, $RMSD_B$ > $RMSD_A$ grey-shaded areas at ≈ 0.7 μs). An RMSD measure modified by a multiplicative term linearly scaled by the distance of an atom to the center of the DNA (Fig. S3) confirms the expected result that the duplex termini are the most dynamical regions (*i.e.*, account for the bulk of the RMSD).

Fine-grained differences between DNA conformations over various time spans of the trajectory were assessed by pairwise RMSD matrices, computed over periods ranging from ns → μs. As shown in Fig. 2 and S4, short-time (ns-scale) matrices capture those features of the trajectory which may be expected, such as the restrained dynamics phase of the equilibration period (first ns) corresponding to very low RMSD values and differing significantly from the remainder of the trajectory (Fig. 2a). The long-time (μs-scale) matrix represents $>10^{12}$ pairwise comparisons (Fig. 2b); significant off-diagonal basins of increased structural similarity/dissimilarity in this matrix illustrate that the notion of an average DNA structure is not an accurate representation of the true ensemble of thermally-accessible conformational states in the μs-scale dynamics of DNA. [A similar point has been made by Beveridge *et al.* (33).] Patterns of intense variability in the μs matrix (*e.g.*, the striping at long times in Fig. 2b) show that much of the structural variability arises only beyond ≈ 0.6-0.7 μs. These patterns also served as initial hallmarks of the large-scale structural transitions which were subsequently identified as a base flipping event (Fig 2b, orange arrow) and a cross-strand intercalative stacking transition (Fig. 2b, yellow arrow).

**Cross-strand intercalative stacking (XSIS)**

An intriguing example of DNA structural plasticity and conformational polymorphism that emerged only at long times involves disruption of a Watson-Crick base pair *via* cross-strand intercalative stacking ("XSIS") of the bases in the previously intact pair. This atypical XSIS state develops into a stable, persistent form between ≈ 0.7–0.75 μs, and occurs at the $(A•T)_{12}$ bp in an A/T-rich region near the center of the κB recognition element (Fig. 1a). The aforementioned patterns of variability in pairwise RMSDs (Fig. 2, S4) suggested the large-scale structural perturbations accompanying the XSIS transition. Visual analysis of the trajectory (*Supp. Movie A*), as well as calculation of bundles of κB DNA conformers and corresponding averaged structures over discrete time windows (Fig. 3), revealed the cross-strand base stacking that characterizes the XSIS state. This transition involves essentially complete abrogation of the $(A•T)_{12}$ pair, such that the constituent bases ($Ade_{1,12}$ and $Thy_{2,9}$) longitudinally shear apart and assume a nearly coaxial rather than coplanar arrangement – *i.e.*, they become stacked upon one another (Fig. 4c-e), with the $A_{1,12}$ staggered towards the 5' direction and the cross-strand partner $T_{2,9}$ translated towards the 3' direction (with respect to the 5'→3' path of the parent strand along the global helical axis). Though fundamentally different from the longitudinal breathing mechanism proposed by Harvey over twenty years ago for B ↔ Z DNA conversion (34), the XSIS transition is similar in spirit, insofar as stretching of the local backbone along the helical axis creates free space to accommodate an additional object (the new 'object' being a longitudinally sheared base in XSIS, versus a propagating cavity which accommodates intact base pairs as they flip their α/β ring faces in the Harvey model). There is also a degree of structural similarity between XSIS and the pattern of inter-strand interactions arising during simulations of duplexes subject to an applied tension, leading to a stretched '*S*'-form DNA (35); however, any relationship between *S*-DNA and XSIS is unclear (*i.e.*, the similarity may be a purely geometric result arising upon lengthening of any double helical arrangement of planar, interacting groups).

The XSIS transition can be monitored by both geometric and physicochemical parameters. Among the standard rigid-body parameters describing local bp and bp step geometry, the intra-bp *Stagger* ($S_z$) is particularly suitable for capturing the XSIS process (Fig. 4a, S6). Likewise, the cross-strand mutual base overlap area described in Fig. 5 (and ref. (36)) measures the degree of inter-strand stacking of the bases participating in XSIS, and can be seen to be a highly sensitive gauge of XSIS-like transitions (including the possibility of discerning asymmetric behavior of the two 5'- and 3'-shearing bases). Disruption of the local helical stack due to XSIS is accompanied by compensating (inversely correlated) changes in the bp and bp step parameters of neighboring nucleotides (Fig. 4a, S6). Although the perturbative effect of this transition does not propagate very far along either direction of the duplex in terms of these geometric parameters (Fig. S6), the timing of the XSIS transition nearly coincides with formation of a barbed terminus (described below), suggesting some degree of dynamical coupling between sites >10 base pairs apart in sequence (Fig. 1a). As may be expected from the geometry of the double helix, local groove width is one of the strongest structural correlates of XSIS (Fig. 4b): The transition coincides with perturbation of the proximal minor groove, wherein an initially widened groove over the ≈ 0.6→0.74-μs time span (relative to the ≈ 5.8 Å width in canonical *B*-form DNA and the even narrower width characteristic of A-tracts) rapidly constricts in the 5' direction. These structural features spur great interest in the possible roles of condensed counterions in triggering and/or modulating XSIS-like transitions.

An initial assessment of potential coupling between the counterion atmosphere and structural transitions such as XSIS was made by calculating number densities of $Na^+$ ions within 15 Å of the κB DNA over the course of the trajectory, with populations being averaged over 1-, 10-, and 100-ns windows centered on the 'early' (Fig. 4c), 'middle' (Fig. 4e), and 'extreme' (Fig. 4f) states of XSIS. These results (Fig. 4d and unpublished data) are consistent with known structural trends – *e.g.*, preferential localization of $Na^+$ density near more strongly electronegative regions of nucleoside bases (Ade N6/7, Thy O2, *etc.*). Consistent with many X-ray, NMR, and MD studies indicating cation-dependent minor groove narrowing (9,11,12), the severely constricted minor groove near the XSIS site (Fig. 4c,e) exhibits the highest amplitude peak in the 10-ns window-averaged $Na^+$ density map (Fig. 4d, right panel). Nevertheless, the detailed linkage between groove-localized $Na^+$ ions and structural transitions such as XSIS remains unclear; given the lengthy timescales over which the XSIS and flipping events evolve (>100 ns from initial onset to final resolution), it is likely that the local ionic environment (relaxing on the ns timescale for monovalent cations) 'responds' to (rather than drives) the large-scale structural transitions of XSIS and flipping.

The XSIS transition and its potential role in κB DNA dynamics is a novel finding attributable to the μs-scale sampling of our trajectory, and it should be emphasized that this simulation-based result is more predictive than conclusive. Solution NMR studies of other κB DNA elements have been pioneered by Hartmann and colleagues (37); however, to our knowledge, MD simulations of a κB DNA have not been performed. Therefore, prior knowledge of what may be anticipated in the μs-scale behavior of this κB DNA is limited, and no *a priori* assumptions were imposed with regards to the computed behavior of this DNA fragment. Simulation caveats notwithstanding, it should be noted that numerous experimental data from both NMR and crystallography support or are consistent with an XSIS-like transition. Observations of severe groove-bending immediately preceding XSIS (*e.g.*, Fig. 4f) are consistent with NMR data showing a pronounced (≈ 20°) bend towards the major groove in a related κB DNA (37). Also, several lines of crystallographic evidence are consistent with the XSIS phenomenon. Early studies of poly(A) and $(CA)_n$ tracts by Klug, Moras, and others (38,39) have revealed a "zipper" pattern of bifurcated hydrogen bonds along the axial stack. These favorable $i_1 \cdots j_2$ and $j_1 \cdots i_2$ hydrogen bond interactions (in the notation of Fig. 5) recapitulate the edge/edge interactions which occur between bases in both the *initial* and *final* stages of XSIS: *initial*, as aromatic base planes begin shearing to the "early" state illustrated in Fig. 4d; and *final*, as the XSIS thymine ($T_{2,9}$) begins hydrogen bonding to the "wrong" cross-strand adenine ($A_{1,13}$) and eventually resolving the XSIS state by leading to flipping of its $T_{2,8}$ neighbor. Independent structural work on G◦A mismatches found a similar pattern of $(i/j)_1 \cdots (j/i)_2$ cross-strand bifurcated hydrogen bonds (40), suggesting such cross-strand interactions as a fundamental feature of DNA deformability. Perhaps most relevantly, Reid and coworkers have amassed a large body of NMR results showing that flanking G◦A mismatches (such as occur in human centromeric $(GGA)_2$ segments) can induce the shearing of intervening bases into a "GA-bracketed G-stack" similar to the XSIS state reported here (see, *e.g.*, (41,42)). These experimental data concur with our prediction of a μs-scale XSIS-like transition at the junction of κB half sites.

**Base flipping at the junction of κB half sites**

The most intriguing and unanticipated result to emerge in the long-time dynamics of κB DNA is a spontaneous base flipping event. Because of its high activation barrier, such a transition is considered a "rare event", thought to occur only infrequently in the equilibrium dynamics of a DNA duplex at room temperature (reviewed in (43)). Indeed, flipping of a thymine base at the junction of κB half sites (Fig. 1a) arose only near the end of the μs-long κB trajectory (Fig. 2b, 3e,f), at ≈ 950 ns as measured by several geometric criteria (Fig. 6, S6; see below). It is interesting that flipping occurred at $(A{\bullet}T)_{13}$, immediately adjacent to the disrupted $(A{\bullet}T)_{12}$ XSIS site. The thymine base that is the focal point of this transition ($T_{2,8}$) can be seen to twist entirely out of the double helical stack *via* the major groove (*Supp. Movie B*), thereby eliminating the $(A{\bullet}T)_{13}$ Watson-Crick bp and yielding an apyrimidinic lesion (Fig. 3e,f, 6b). Concomitant closure of the newly-formed lesion by pairing of the 3' neighboring thymine ($T_{2,9}$; of the XSIS pair) to the orphan adenine ($A_{1,13}$) suggests the model of XSIS-facilitated flipping presented below.

The flipping transition can be monitored by the time-evolution of *(i)* local bp and bp step parameters; *(ii)* correlated changes in local helical geometry (*e.g.*, minor groove structure); and *(iii)* physicochemical quantities such as base overlap areas and the solvent-accessible surface area of the flipping base. Of the standard parameters used to describe bp and bp step geometry (13), the *Opening* ($\sigma$) angle is a useful descriptor of the flipping transition (Fig. 6a). [Note that $\sigma$ is not the same as the angular parameters developed as reaction coordinates in landmark studies of forced flipping by the groups of Lavery (an inclination-independent construction accounting for local helical geometry; see (44)) and MacKerell (a center-of-mass pseudodihedral; see (45)).] The distance between complementary bases is also a useful descriptor of the flipping transition ($d_{COG}$, Fig. 6b), with the advantage that it is independent of the relative angular orientation between base-centered reference frames. As is the case with XSIS, significant redundancy and correlation exists within two distinct types of data: *(i)* The time-evolution of the various other rigid-body bp and bp step parameters are naturally coupled to $\sigma$ and also serve (to varying degrees) as reporters of the flipping event; for example, the translational *Slide* ($D_y$) and angular *Tilt* ($\tau$) of a bp step are rather sensitive to flipping, whereas the *Roll* ($\rho$) is less so (Fig. S6). *(ii)* Time series of a given parameter across the different base pairs and base pair steps near the XSIS $(A{\bullet}T)_{12}$ and flipping $(A{\bullet}T)_{13}$ sites are correlated (see, for instance, the flanking $(A{\bullet}T)_{11}$ bp lying 5' to the XSIS site; Fig. 4–6, S6). In terms of local helical geometry, an intriguing negative correlation exists between the dynamics of the minor groove widths at bp steps bracketing the XSIS and flipping sites – $(AA/TT)_{11}$ and $(AA/TT)_{12}$ on the 5' side and $(TT/AA)_{14}$ on the 3' side (with respect to *strand-1*; Fig. S1). The somewhat narrowed minor groove on the 3' side of the flipped $T_{2,8}$ (suggesting the *B'* form of DNA) significantly widens (Fig. 4b, blue trace), whereas the significantly widened minor groove on the 5' side of the flipping/XSIS site drastically narrows (Fig. 4b, yellow and red). The crossover point between these minor groove transitions (Fig. 4b, ≈ 750 ns) marks the onset of XSIS; the subsequent flipping event is evidenced by a further (smaller-scale) perturbation in the 3' groove widths (Fig. 4b, blue and green traces at ≈ 950 ns).

Comparison of the μs-scale κB DNA flipping event to the vast literature on both experimental (15) and computational (43) studies of base flipping is hampered by the single observation of this transition, as well as the fact that, to our knowledge, this is the first suggestion for a possible role for base flipping in κB DNA dynamics (*i.e.*, κB flipping data do not exist). Much of the existing flipping work focuses upon protein-facilitated ("activated") flipping, rather than the spontaneous ("passive") flipping reported here (15). However, numerous experiments have established that spontaneous DNA and RNA base flipping occurs, the methods ranging from chemical approaches (such as trapping a transiently flipped base in a macrocyclic host molecule (46)) to NMR-based measurements of imino proton exchange (47,48). On a related note, Dickerson and coworkers discovered a novel instance of intermolecular intercalation in the crystal structure of a hybrid DNA/RNA duplex (49), an implication being that such "base pair swapping" occurs *via* spontaneous base flipping. Most notably, very recent crystallographic work has revealed a base extruded from the stack of a related HIV-1 κB DNA element, corroborating our prediction that base flipping may occur in κB DNA (50).

Discovery of κB DNA base flipping in the μs regime is a fortuitous result, as NMR studies of other DNA sequences find that this process occurs on a longer timescale (ms), inaccessible to equilibrium MD simulation. Nonetheless, that the base which flips is a thymine is consistent with the relative pyrimidine versus purine energy barrier, as well as with NMR exchange data revealing that *(i)* the lifetime of an A•T bp is generally shorter than G•C pairs (47) and *(ii)* spontaneous A•T opening is the basis for recognition of extrahelical bases (Ura/Thy) in the uracil glycosylase DNA repair system (51). (As discussed below, the junction of half sites in κB elements is almost always an A•T bp.) Also, with regards to characteristic timescales for flipping – and the base pair opening or "breathing" which necessarily precedes it – previous simulation work has found spontaneous breathing on the ns timescale for a difluorotoluene-substituted A•F pair (52). Flipping of $T_{2,8}$ *via* the major groove is consistent with both experimental studies (53) and recent theoretical work (45,54,55) on flipping pathways. Strong coupling between base flipping and helical axis bending was found in early theoretical work (56) as well as more recent computational studies (55). Consistent with those findings and the aforementioned NMR studies (37), the κB DNA exhibits significant helical axis bending in a most extreme state of XSIS preceding the base flipping event (Fig. 4f).

Finally, we note that the κB flipping reported here is not necessarily inconsistent with NMR-based findings of generally slower A•T opening in A-tracts (47). As with any computational or experimental approach, the NMR exchange methodology rests upon various assumptions and approximations. In particular, known limitations and caveats of this technique include *(i)* assumptions of two-state behavior (entirely open/closed bps); *(ii)* ambiguity of the actual physical process being detected (accessibility of exchanged protons versus complete base extrusion); *(iii)* the neglected structural and dynamical influence of ions (such as the ammonia exchange catalyst used in these studies) on local DNA structure (57); and *(iv)* intrinsic detection limits (ms timescales being the lower bound). Similar NMR studies have shown that A•T opening strongly depends on sequence context, and is also sensitive to oligonucleotide length (58). Most significantly, the structural basis of greater A•T lifetimes in A-tracts is assumed to stem from a presumably narrower (and more rigid) minor groove in these tracts ((43) and references therein). However, the minor groove of the κB element does not exhibit this behavior immediately prior to the XSIS → flipping cascade (Figs. 4b, S6), and therefore does not adhere to this assumption. It should be reiterated that ultimate assessment of predicted κB base flipping will require future experimental studies.

**XSIS-facilitated flipping?**

XSIS and base flipping appear to be intimately linked. Coupling between XSIS at $(A{\bullet}T)_{12}$ and base flipping at the 3' neighboring $(A{\cdot}T)_{13}$ is reflected in the time-evolution of numerous structural parameters between ≈ 750 ns (onset of XSIS) and 950 ns (onset of flipping). This particular 200-ns period is immediately preceded by perturbation of local helical structure, as quantified by, *e.g.*, minor groove widths in the vicinity of these bp steps. We suggest that cross-strand intercalative stacking may facilitate passive, spontaneous base flipping – *i.e.*, flipping in the absence of proteins that can actively extrude, bind, and stabilize flipped bases (*e.g.*, DNA methyltransferases (15,59) and glycosylase repair enzymes (60)). Such a mechanism may be especially true in the case of the particular A/T-rich region at the center of the κB DNA (Fig. 1), as this segment is highly degenerate, of low sequence complexity, and is therefore amenable to "strand-slippage" deformations (61,62). In this model, the central Ade (*n*) in an $\cdots AA_nA\cdots$ tract undergoes a thermally-induced, transient breathing event (52,63), staggers into an XSIS-like state (quantified by $S_z$), and then, in re-annealing to the complementary strand by re-forming the A•T pair, does so with a neighboring ($n \pm 1$) thymine instead of the original cross-strand partner. Evidence for such a mechanism is provided by static snapshots (Fig. 4, 6) and averaged structures (Fig. 3, S4), and most clearly by visual inspection of the process as it occurs (see *Supp. Movies A, B, D*). Note that an A-rich region of the low-complexity ···AAA··· form found at the junction of κB half sites (rather than a different permutation across the strands, such as ···ATAT···) is perhaps *the* ideal sequence on which such a mechanism could act, as it would be more amenable to such $n \rightarrow n \pm 1$ slippage than other configurations of base pairs.

Evaluation of our model for XSIS-facilitated flipping in DNA (κB or otherwise) ultimately requires experimental data. Specific studies using experimental and/or computational techniques can be envisaged as being particularly well-suited for exploring the XSIS transition, the base flipping event, and the key coupling between these events that we posit as a mechanism for spontaneous base flipping in A-rich tracts. Flipping in the long-time dynamics of this κB DNA could be studied *via* the chemical (flipped base trapping) and biophysical (proton exchange) methods outlined above, as well as a recently-developed selective, non-covalent base flipping assay (64). Simulation-derived dynamics of this κB element also could be tested in terms of agreement (on the ≈ 50 ns timescale) with experiments using the time-resolved Stokes shift spectroscopic method that has been fruitfully applied to DNA by Berg and coworkers (65). Perhaps most compelling, the XSIS transition (and coupled flipping) could be explored by assaying the charge transfer properties of DNA duplexes containing this and related κB sequences. Conformationally-gated charge transfer through DNA has been established by Barton and colleagues as a sensitive gauge of local DNA structure (recently reviewed in (66)), and the sequence-sensitivity of this phenomena has been demonstrated (67). Surveying the anticipated disruptive effect of an XSIS-like state (see Fig. 4e,f) upon DNA charge transfer efficiency in multiple sequence contexts would serve as a test of the predictive results from our simulation, and the data from such studies would help support or refute a mechanistic model for spontaneous base flipping *via* XSIS-facilitated base pair opening.

**A minor groove-bound "barbed" structure**

DNA oligonucleotide structures in which a terminal nucleoside loops back upon its parent strand to form a stable, well-defined, yet alternative (non-canonical) structure are uncommon: Formation of DNA secondary and tertiary structure is a highly cooperative process wherein complementary strands reproducibly "zip-up" into the same configuration of energetically-stable Watson-Crick base pairs (3). Thus, a κB DNA conformation in which a 3' terminal nucleotide "peels" away from the helical stack and forms favorable (hydrogen bonded) interactions with the proximal minor groove of its parent strand was unexpected (Fig. 3c-f, 7, S7). It is intriguing that this barbing occurs at the $(G{\equiv}C)_1$ rather than the $(A{=}T)_{20}$ terminus of the non-palindromic κB DNA duplex (Fig. 7), given that solvent-exposed triply hydrogen-bonded G•C pairs are more stable than A•T pairs. As a positive control on the dynamical behavior of the κB DNA trajectory with respect to the termini, the large-scale pattern of fraying at the $(A{\bullet}T)_{20}$ terminus (Fig. S7) and the train of transient fraying events at $(G{\bullet}C)_1$ prior to barbing (Fig. 7) are consistent with the thermally-induced terminal fraying previously observed in ns-scale MD simulations of nucleic acids (68). Neglecting intermolecular terminus···groove contacts in the lattices of some DNA crystals, the most structurally similar examples to barbing of which we are aware are the minor groove association of extra-helical cytosines in the solution structure of a crosslinked DNA ((69); the Cyt base is not near a terminus), and the "G1" conformation found in simulations of a DNA octamer containing an adenine bulge (the bulged adenine occasionally makes edge···minor groove contacts, (70)).

In addition to being structurally (Fig. 7, inset) and dynamically distinct from the known stochastic fraying of DNA termini, the κB DNA barbed terminus emerges only at very long times (appearing at ≈ 750 ns, it is likely to be coupled to XSIS), and can stably persist for periods on the order of 100 ns. The fact that the barbed state reproducibly and independently arises over the course of the same trajectory (multiple ≈ 7.5-Å plateaus in the blue trace of Fig. 7) bolsters its potential significance, although the biological significance of this DNA conformation is inherently limited by it being an "end effect." Nonetheless, a barbed-like structure may be relevant from other perspectives, including the design of minor groove-binding agents (71), and in light of the numerous biochemical assays that rely upon the use of oligonucleotides not unlike the κB DNA element studied here. The barbed κB terminus suggests that addition of even up to three G•C pairs (Fig. 1) to reinforce a presumably weak terminus does not suffice to prevent such end-effects from occurring on the μs timescale.

### Validation of the trajectory and DNA backbone dynamics on the μs timescale

The present work on κB DNA focuses on the unexpected XSIS and base flipping events rather than a comprehensive analysis of the μs-scale conformational dynamics of the DNA backbone or technical aspects of MD force field development. Nevertheless, methodological studies of force field shortcomings and parameterization are of immense concern and relevance in biomolecular MD simulations, and are the subject of much active investigation ((72) and references therein); our current simulation provides additional data for such studies, as well as a host of definite, testable predictions about a specific μs-scale dynamical process that may be investigated by both theoretical and experimental means. Compared to NMR and crystallographic data, existing force fields are known to yield discrepancies between trajectory-averaged values of base pair and base pair step parameters such as the *Roll* (ρ) and *Twist* (ω; (73)). In particular, MD-derived values of *Roll* are generally moderate overestimates ($\rho_{MD} \approx 4$–5°, $\rho_{NMR} \approx 3°$, $\rho_{x\text{-}ray} \approx 0°$), while simulations typically underestimate the average *Twist* ($\omega_{MD} \approx 30°$, $\omega_{NMR/X\text{-}ray} \approx 34°$). The μs-scale DNA trajectory reported here recapitulates these trends, both within the XSIS/flipping region (Fig. S6d) and at a distal site that effectively serves as an internal control for the behavior of these parameters (Fig. S6g). On a larger structural scale, two aspects of backbone dynamics identified as being particularly relevant based on ns-scale MD simulations are transitions between the $B_I/B_{II}$ substates (74) and possibly insufficient (and/or imbalanced) sampling of long-lived backbone states arising from concerted (crankshaft) α/γ rotations (30,72,75). For instance, a concern in ns-scale DNA simulations is that the trajectory may improperly sample (or even become indefinitely trapped in) non-canonical regions of α/γ conformational space, leading to force field-dependent artifacts.

The α/γ and $B_I/B_{II}$ sampling properties of the trajectory were examined, both to gain an initial understanding of the backbone dynamics of this κB DNA on the μs timescale, and also to gauge possible artefactual sampling due to force field imbalances (*e.g.*, becoming trapped at non-canonical α/γ values for half of the trajectory). Two distinct (but interrelated) facets of the α/γ sampling issue involve *(i)* the relative frequencies of visiting various α/γ states (*i.e.*, free energy differences), such as the known $\alpha/\gamma \approx g^-/g^+ \approx (300°, 30°)$ energy minimum; and *(ii)* whether or not the backbone is able to interconvert between distinct (and possibly long-lived) α/γ states, including the canonical $g^-/g^+$ ground state in addition to $g^+/t$ and $g^-/t$ local minima (the latter being a 'γ' flip away from the global minimum). With regards to *(i)*, it is reassuring that it is the $g^-/g^+$ ground state which is occupied by the only nucleotide to not undergo a single α/γ transition over the course of the entire μs (Fig. 8a; interestingly, this is the XSIS adenine). With regards to *(ii)*, the κB DNA trajectory shows extensive sampling of discrete basins of the α/γ conformational landscape on the μs timescale, including reversible transitions to/from the $g^-/g^+$ energy minimum (Fig. 8b). Indeed, the most densely-populated region of this space is the known $g^-/g^+$ ground state, followed by the two regions ($g^-/t$ and $g^+/t$) previously identified as local minima (75). Similarly, the $B_I \leftrightarrow B_{II}$ sampling behavior of the μs-scale trajectory is consistent with existing principles. An angular histogram of the $B_I/B_{II}$ torsional parameter ($\varepsilon - \zeta$) for a nucleotide adjacent to the XSIS position reveals extensive sampling near the favorable $B_I$ state that is found in canonical *B*-form DNA ($\varepsilon - \zeta \approx -90°$), in addition to a smaller component corresponding to the slightly less favorable $B_{II}$ state ($\varepsilon - \zeta > 0°$). The complete body of α/γ and $B_I/B_{II}$ sampling data (*i.e.*, for all 38 nucleotides) indicates that the κB DNA trajectory reasonably samples these backbone substates, and does not exhibit artefactual "trapping" in anomalous, high-energy states. Several features of κB DNA backbone dynamics are in agreement with a recent μs-scale dodecamer simulation (31), including the differential features of α/γ and $B_I/B_{II}$ time series sampling for A·T and G·C bps (data not shown).

The overall μs-scale structural integrity of the DNA duplex, as well as any systematic biases towards a particular helical morphology (ideal *A*-, *B*-, *etc.* forms) due to force field-specific limitations, were assessed by classifying the DNA conformation as either *A*, *B*, or *TA*-like at each timepoint. Such analysis of the μs-scale trajectory in terms of the sampling of known DNA helical structures provides a form of simulation validation that lies beyond the local level of nucleotides and base pairs. The phosphate-based $z_P/z_P(h)$ metric has been shown to be a particularly effective discriminator between the *A*, *B*, and *TA*-like forms of DNA (76). Details of this quantity as a classifier of DNA structure are provided in Figs. 9 and S8 (see the arrows denoting *A*-, *B*-, and *TA*-like regions). Note that the most significantly perturbed helical steps in the XSIS and flipping events ($(AA/TT)_{12}$ and $(AT/AT)_{13}$) are also the ones which exhibit multiple discrete clusters (green patches in Fig. 9b, purple in Fig. S8a); nevertheless, the preponderance of the trajectory is spent in the *B*-like regions of $z_P/z_P(h)$ space for each step of the κB duplex (data not shown). These scatter plots demonstrate that the overall conformation of the duplex – *i.e.*, its structure at a slightly more global level than that addressed by geometric quantities such as local bp parameters and backbone torsion angles – is well-maintained throughout the μs-long simulation, even at highly dynamical sites that exhibit the most significant structural variability (*i.e.*, XSIS and flipping).

Also in connection with potential force field-induced artefactual behavior, note that the starting DNA structure was somewhat "distorted" in the region distal to the κB site – *i.e.*, it was not perfectly identical to canonical *B*-form DNA (see, *e.g.*, Fig. 3a). However, it was found that the simulation effectively "regularized" this region of the DNA over a broad time span (≈ 350-600 ns), yielding a duplex structure more similar to canonical *B*-form DNA than was the starting structure (see, *e.g.*, the 1→100 ns bundle in Fig. 3b and the depressed plateau in the RMSD traces of Fig. S3). This is a significant observation, as it effectively serves as an internal control of simulation quality and demonstrates that the trajectory was neither driven directly into a *B*-form helix by the force field, nor did it monotonically degrade into an artefact-ridden state capable of producing transitions such as XSIS and base flipping.

**Conclusions, and implications for NF-κB···DNA recognition**

The XSIS and base flipping events spur great interest in potential links between these transitions and the exact mechanism(s) of NF-κB···κB DNA recognition. What dictates the specificity of NF-κB···DNA binding? A partial answer comes from recognition of the modular nature of NF-κB–bound κB elements:

Crystal structures (*e.g.*, (21)) show that one NF-κB monomer typically forms most of the contacts to one κB half site (*e.g.*, the 5' AGAA in Fig. 1a), while the other subunit primarily contacts the other half site. This implies a model of "cognate site recognition" (21), wherein the κB DNA specificity of a given NF-κB dimer is a composite of individual half site preferences. However, numerous exceptions to this model exist (the consensus sequence is weak), and NF-κB is somewhat atypical in its interactions with DNA: *(i)* DNA-contacting side chains reside in the loops of NF-κB (rather than secondary structure elements); *(ii)* many DNA contacts are mediated by water molecules (rather than direct side chain···base interactions; see, *e.g.*, (77)); and *(iii)* a large fraction of DNA contacts are to the phosphodiester backbone (rather than the sequence-specific grooves). Together with the biochemical and biophysical studies on which they are based, these principles suggest that the mechanism by which a particular NF-κB dimer recognizes and binds to a specific κB site is one of exquisitely subtle indirect readout.

A mechanistic basis for this readout is implied by the μs-scale conformational properties of the κB DNA element studied here: The cognate site recognition model should be broadened to account for the dynamical behavior of target κB sites, including the possibility of transient base flipping. The occurrence of XSIS-facilitated flipping at the junction ($N_0$) of κB half sites in the $A_{-4}G_{-3}A_{-2}A_{-1}N_0T_{+1}T_{+2}C_{+3}C_{+4}$ element is extremely salient for several reasons. First, several features of known κB sequences are consistent with the model proposed above for XSIS-based flipping, including the fact that the central positions ($-2 \rightarrow +2$) of naturally-occurring κB elements are generally A/T-rich, in organisms ranging from insect (78) to human (79). Also, *in vitro* selection experiments for sequences which optimally bind to different NF-κBs recapitulate this trend, frequently finding a consecutive run of A/T's near the middle of the optimal κB DNA sequence (80). Finally, systematic surveys of κB sequence preferences *via* both array-based experiments and informatics analyses have revealed similar patterns of κB elements with A-rich centers; most intriguingly, these analyses also found strong coupling between base pair positions 0 and +1 (*i.e.*, where the flipping occurs in our simulation).

Detailed conformational and dynamical properties of κB DNA sequences likely modulate the binding of NF-κB dimers to various κB sites, as noted in crystallographic studies of an unbound class II κB DNA (81). A similar perspective on the important role of dynamics originates in NMR studies of HIV κB DNA (37). More generally, sequence-specific DNA curvature and flexibility have been suggested as essential in modulating affinity and specificity in other transcription factor systems, including, for example, the specific recognition profiles of TATA box-binding proteins for potential high-affinity DNA sites (82). Expanding upon this idea, a primary and experimentally-testable prediction of our μs-scale MD studies is that base flipping may occur in κB DNA systems, with a transiently extruded thymine playing a key role in indirect readout of a dynamical κB recognition motif. Exploration of this possibility will likely deepen our understanding of the malleability of the DNA double helix, as well as the significance of such flexibility in mediating protein···DNA recognition in the NF-κB gene regulatory network.

## Supplementary Information

The following supporting material is available: *(i)* Methodological details and associated references; *(ii)* Additional figures and accompanying captions; and *(iii)* Video animations illustrating the XSIS, flipping, and barbing transitions, as well as an overview of the full μs-long trajectory.

## Acknowledgements

We thank L. Chen, L. Columbus, J. Dzubiella, J. Feigon, G. Ghosh, J. Gullingsrud, D. Hamelberg, P. Qin, J. Trylska, and L.D. Williams for helpful discussions, and R. Konecny (UCSD) for exceptional computer support. We gratefully acknowledge the Howard Hughes Medical Institute (HHMI), National Biomedical Computational Resource (NBCR), National Institutes of Health (NIH), National Science Foundation (NSF), and the NSF-funded Center for Theoretical Biological Physics for financial support (J.A.M.), as well as funding from a Sloan/DOE postdoctoral fellowship in the initial stages of this work (C.M.).

## Figure Legends

**Fig. 1. Overview of the κB DNA element and simulation system.** The μs-scale dynamics of the κB DNA element shown in this sequence schematic (**a**) were explored, the MD simulation system consisting of the 20-base pair duplex immersed in a bath of explicit water and 50 mM NaCl (**b**; dark blue line indicates perimeter of cut-away frontal slice). This particular κB DNA sequence is bound by the $(c\text{-}Rel)_2$ NF-κB homodimer, and is essentially a composite recognition element consisting of AGAA and TTCC κB "half sites". Much of the interesting structural and dynamical behavior of this A/T-rich duplex (A-rich regions are accentuated in red) arose within the nonameric κB element (grey background in **a**, CPK spheres in **b**), including the cross-strand intercalative stacking (XSIS) and base flipping events.

**Fig. 2. Conformational plasticity of the κB DNA duplex on the ns and μs timescales, as assessed by pairwise RMSD matrices.** Matrices of pairwise coordinate RMSDs are shown for the first 10 ns of the trajectory (**a**) and over the course of the entire μs-long production run (**b**). The first and last timesteps are indicated (*e.g.*, in **b**, $t_1$ is the structure after exactly 1.0 ns of equilibration), as are trajectory sampling frequencies ($\delta t$) and the min/max RMSD values used for linearly-scaled coloring of matrix elements (color bars at right). Features of the heating and equilibration strategy are evident in the 10 ns matrix, including the imposition of restraints over the first ~ 400 ps (dark region in **a**), as well as subsequent drift towards a stable, equilibrated structure (intense yellow stripe); the submatrix corresponding to this initial 1-ns equilibration period is delimited by green lines.

**Fig. 3. DNA polymorphism: A 'barbed' terminus, cross-strand intercalative stacking ('XSIS'), and spontaneous base flipping.** The XSIS, base flipping, and barbed terminus transitions are illustrated in these bundles of κB DNA conformers, superimposed to a common reference structure over time spans ranging from the initial 1-ns equilibration period (**a**) and ensuing 100 ns of dynamics (**b**) to the final 320 ns of the 1,021-ns trajectory (**c**→**f**). Snapshots of the DNA backbone within each bundle are indicated as thin lines (equally spaced in time), and progress within each time series is indicated by grading of backbone colors from red (first) → grey → blue (last). Locally-averaged structures were computed (stick representation for non-hydrogen atoms and tan-colored backbone), as were the mean global helical axes (green spline); for clarity, base and deoxyribose rings are colored red (strand-1) and blue (strand-2), and bases of the 9-bp κB element are rendered as spheres. Regions of greatest structural perturbation are also illustrated as space-filling spheres: *(i)* the cytosine of the "barbed" $(G{\cdot}C)_1$ terminus (lavender); *(ii)* the sheared $(A{\cdot}T)_{12}$ which is the focal point of the XSIS transition (yellow); and *(iii)* the neighboring $(A{\cdot}T)_{13}$ thymine, which is extruded from the helical stack in the base flipping event (orange). Note that the persistence of the perturbed positions of the barbed ($C_{2,20}$), flipped ($T_{2,8}$), and staggered XSIS ($(A{\cdot}T)_{12}$) bases at these sites – even in structures averaged over 100-ns blocks of time – indicates that these are stable, long-lived conformational states rather than transient, metastable intermediates in the DNA dynamics.

**Fig. 4. Time-evolution and structural correlates of the XSIS transition.** Time series of the base pair *Stagger* parameter (**a**) and minor groove widths (**b**) capture the structural changes accompanying XSIS. The *Stagger* ($S_z$) quantifies axial displacement of complementary bases with respect to the local bp reference frame, and therefore serves as an ideal descriptor of the XSIS transition; correlations between the XSIS bp ($(A{\bullet}T)_{12}$; yellow) and the 5' neighboring pair ($(A{\bullet}T)_{11}$; red) are evident in these $S_z$ time series (**a**). Similarly, correlated disruptions of local groove geometry are evident in the trajectory of minor groove widths for bp steps near the XSIS site (**b**). Structures of the κB DNA at various timepoints beyond ≈ 750 ns illustrate (**c**) the onset of XSIS, (**e**) a fully-developed XSIS state, and (**f**) a (relatively infrequent) state exhibiting an extreme degree of XSIS and corresponding to a highly bent helical geometry. Local $Na^+$ counterion populations are shown as semi-transparent density isosurfaces (**d**), averaged over either 1-ns (light green) or 10-ns (dark green) windows; the view into the major groove in the left half of (**d**) is identical to that in (**c**), with CPK spheres omitted for clarity. Conventions used in this and subsequent time series plots include: data are plotted in scatter form for only the latter half of the trajectory (sampled

every 5 ps) and only within ±1 nucleotides of the sites of XSIS (yellow) and flipping (green); locally-averaged values (5-ns window) are drawn as solid lines; coloring and numbering conventions are provided in a topmost sequence schematic (bp steps are indicated by semicircular arcs); A/T-rich regions are accentuated by red coloring; and the κB element is demarcated in grey.

**Fig. 5: Cross-strand base overlap areas as a gauge of XSIS.** The geometric area of overlap between the planes of nearest-neighbor bases provides a measure of the extent of stacking (*e.g.*, aromatic π···π interactions). Of the four distinct intra- and inter-strand combinations for a base-paired dinucleotide step ($i_1/i_2$, $j_1/j_2$, $i_1/j_2$, $j_1/i_2$; see schematic inset), the diagonal terms serve as a particularly sensitive gauge of the degree of cross-strand stacking (*i.e.*, XSIS). The $i_1/j_2$ (**a**) and $j_1/i_2$ (**b**) values are nearly zero for a canonical (non-XSIS) step distal to the XSIS site [$(GT/AC)_3$; grey traces], affording an internal control over the latter 0.5 μs of the trajectory. Raw values of the overlap area are shown as points (color-coded as in the legend), with 5-ns window-averaged values drawn as solid lines and local maxima (within a 5-ns neighborhood) indicated as step patterns. Note that both cross-strand terms capture the development of XSIS as an abrupt rise at ≈ 700 ns, with the reciprocity broken by a greater sensitivity of the $i_1/j_2$ term (**b**) to structural perturbations in the 600-700 ns range preceding XSIS.

**Fig. 6. Correlated disruptions at the junction of κB half sites: Resolution of XSIS *via* spontaneous base flipping.** The spontaneous $(A{\circ}T)_{13}$ base flipping event at roughly 950 ns (≈ 200 ns after the onset of XSIS) can be characterized by the standard base pair *Opening* (σ) parameter (**a**), as well as by the distance between the complementary bases in a given pair (**b**). The latter measure ($d_{COG}$) is taken as the distance between the centers of geometry of the bases, and is calculated from the *Shear*, *Stretch*, and *Stagger* as $\sqrt{S_x^2+S_y^2+S_z^2}$ . A sample structure of the flipped state is shown (**b**; inset), with the distance between the extrahelical base ($T_{2,8}$) and its cross-strand partner ($A_{1,13}$) indicated as a dashed magenta-colored line. Other diagrammatic and graphical conventions are as in Figs. 3, 4.

**Fig. 7. A minor groove-bound 'barbed' terminus.** A bp parameter suitable for monitoring the minor groove-bound barbed terminus is the translational *Stretch* ($S_y$) parameter plotted here (same diagrammatic conventions as elsewhere). $S_y$ values for a bp distal to the barbed transition ($(T{\circ}A)_4$) are also plotted, as a representative example of a stable, unperturbed Watson-Crick bp (*i.e.*, an internal positive control). A representative structure from the first occurrence of the barbing transition (between 740→850-ns) is shown. Note that the barbed transition can be detected as an initial shift in $S_y$ away from the near-zero value of an ideal Watson-Crick pair, and the structural and dynamical stability of this state (versus, *e.g.*, stochastic fraying events) can be inferred from persistence of the new $S_y$ value at a constant value over extended lengths of time (the ≈ 7.5 Å plateau in the blue trace).

**Fig. 8. DNA backbone dynamics on the μs timescale: Extensive sampling of α/γ and $B_I/B_{II}$ conformational substates.** Values of backbone α/γ torsion angles are shown as scatter plots for the two indicated nucleotides, color-coded by progress along the μs-scale trajectory (color bar). The backbone at $A_{1,12}$ (**a**) persists in the energetically-favorable α/γ ≈ $g^-/g^+$ basin for the full length of the trajectory, while multiple transitions between this ground state and a higher-energy (but locally minimal) $g^+/t$ state can be seen to occur at $G_{1,19}$ (**b**). An angular histogram of $B_I/B_{II}$ torsional parameter values (**c**) at the $A_{1,11}$ site (proximal to XSIS) shows preferential sampling of the canonical $B_I$ state (ε – ζ ≈ –90º); the bimodal distribution also reveals a minute population of $B_{II}$ state (ε – ζ > 0º; green arrow).

**Fig. 9. DNA backbone dynamics at the junction of κB half sites: Equilibrium sampling of canonical *A*-, *B*-, and *TA*-like helical morphologies.** Values of the phosphate-based $z_P/z_P(h)$ metric are shown for four base pair steps (11→14, **a→d**) which overlap the XSIS/flipping locus between κB half sites. As in Fig. 8, data over the full μs are plotted as temporally-colored points. The maximal extent of sampling is illustrated by the convex hull (orange line), and regions characteristic of *A*, *B*, and *TA*-like DNA (arrows in **a**) are demarcated by dashed cyan lines.

## Mura & McCammon, Figure 1

a) *strand 1* →

(c-Rel)$_2$ NF-κB–binding element

5′ G G G T T T A A A G A A **A** A T T C C A G A 3′

C C C A A A T T T C T T **T T** A A G G T C T

"barbed" terminus (Figs. 3, 6, S5, S7)

XSIS of bases in $(A \bullet T)_{12}$ (Figs. 3, 4, S5)

base flipping at $Thy_{2.8}$ (Figs. 3, 5, S5)

b)

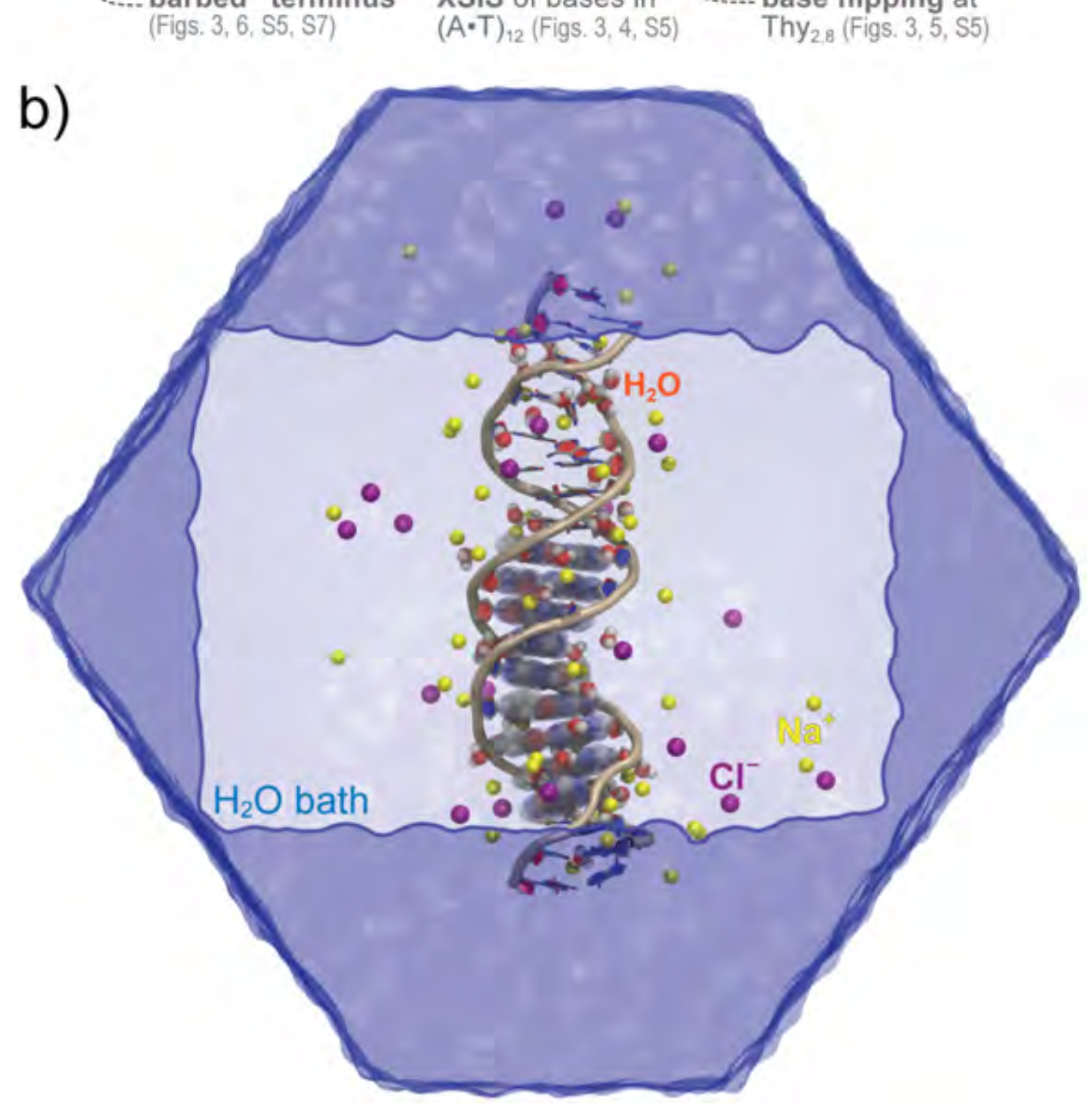

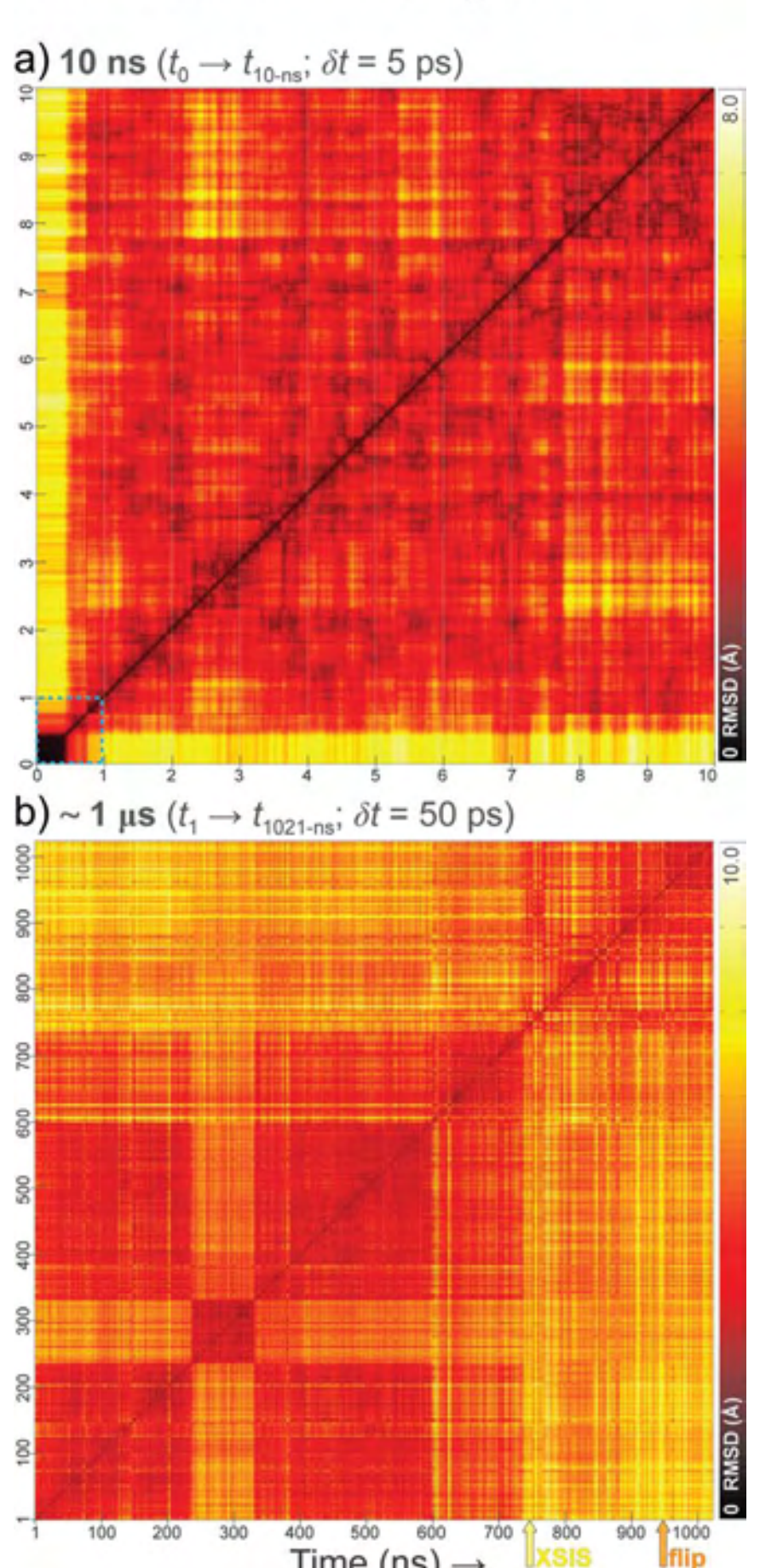

## Mura & McCammon, Figure 3

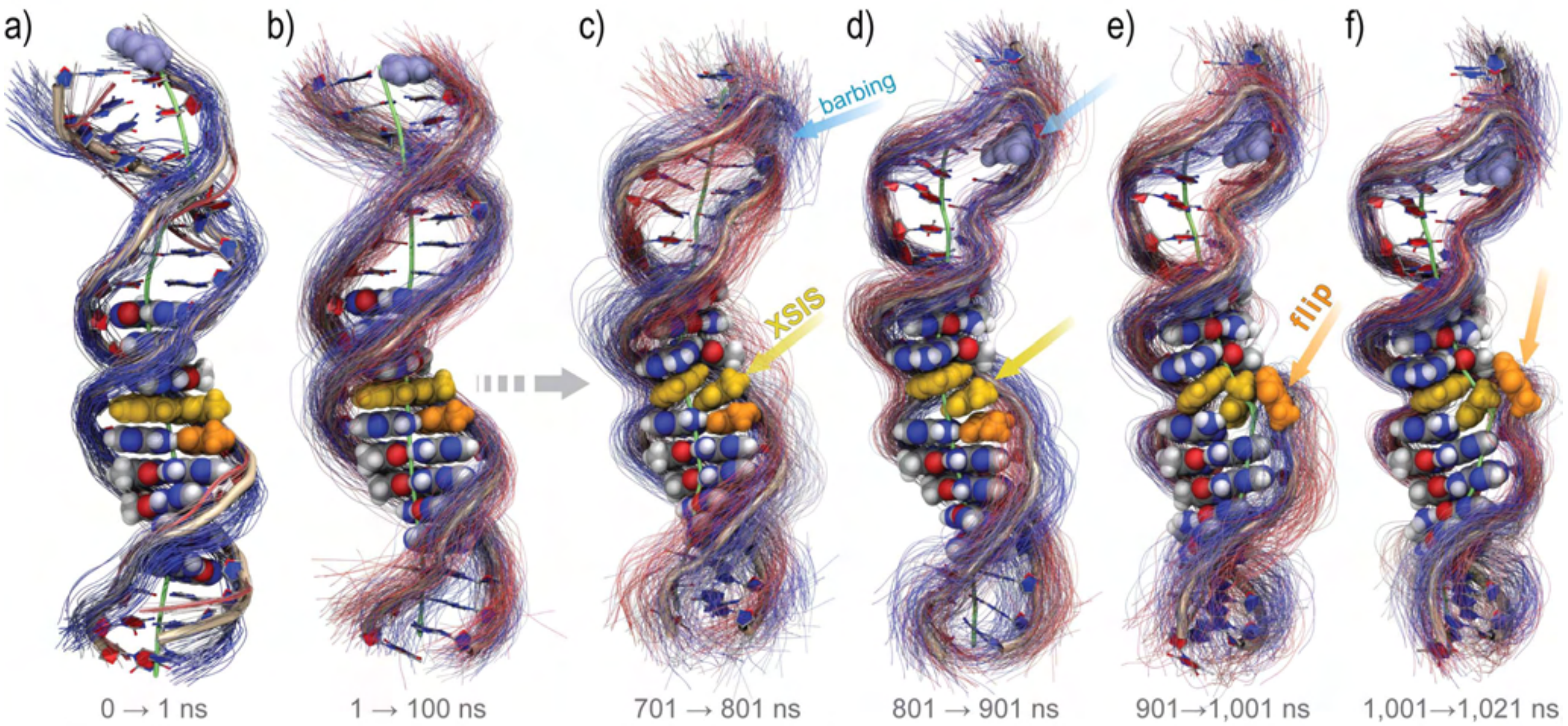

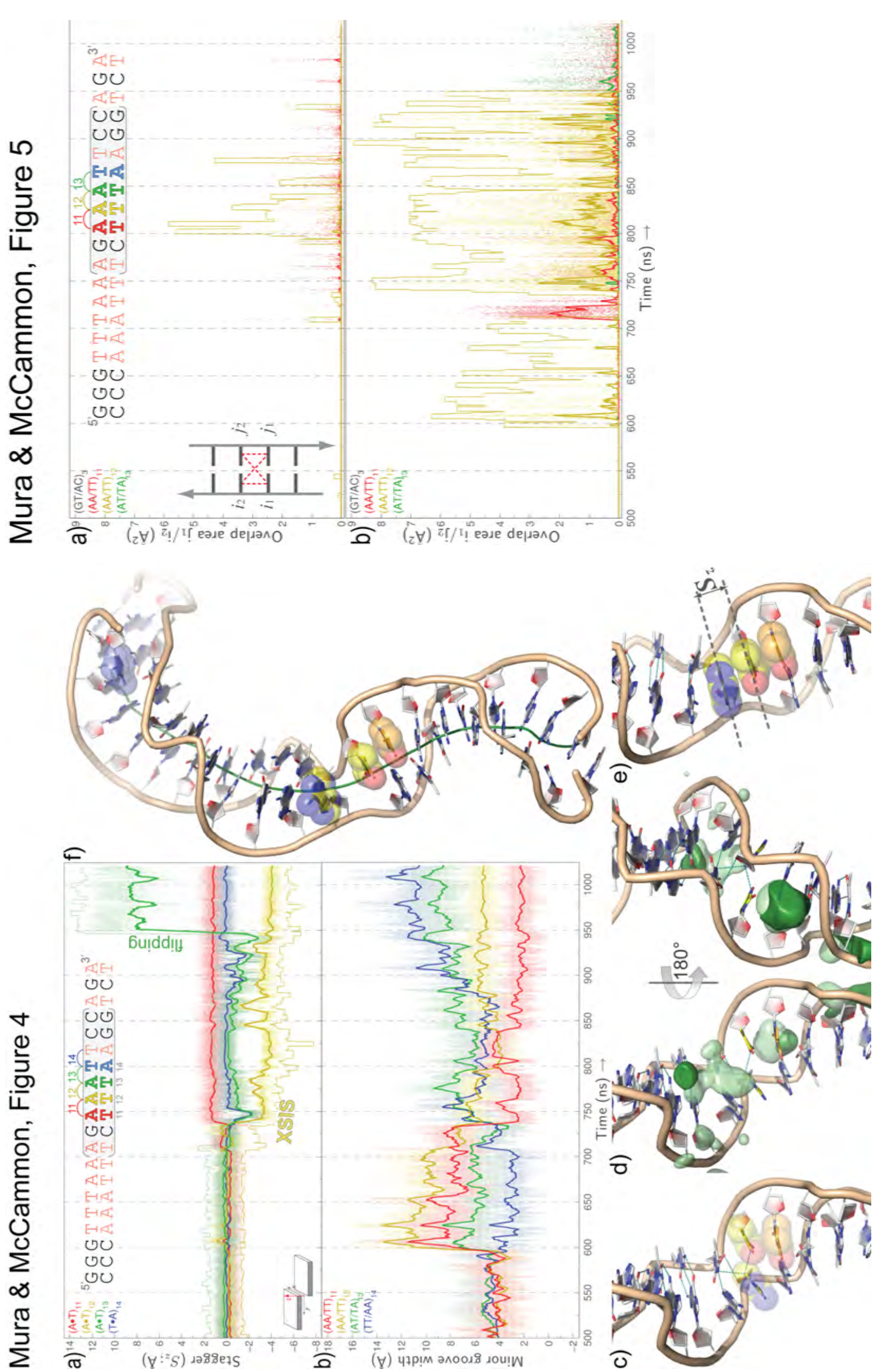

## Mura & McCammon, Figure 6

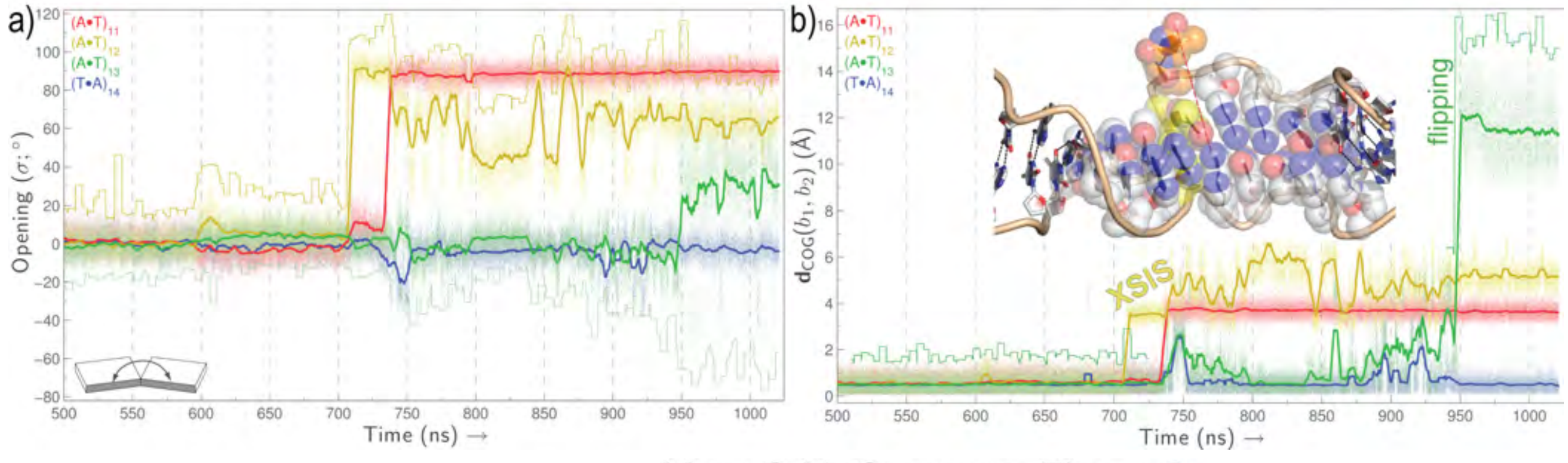

## Mura & McCammon, Figure 7

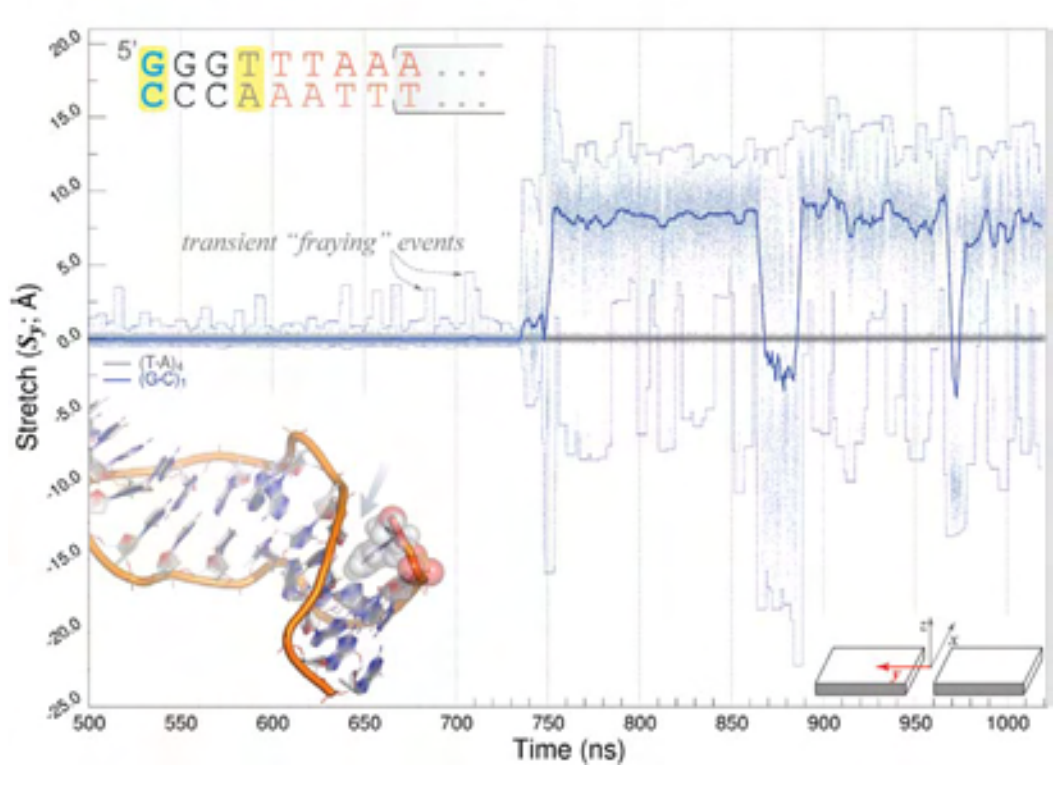

## Mura & McCammon, Figure 9

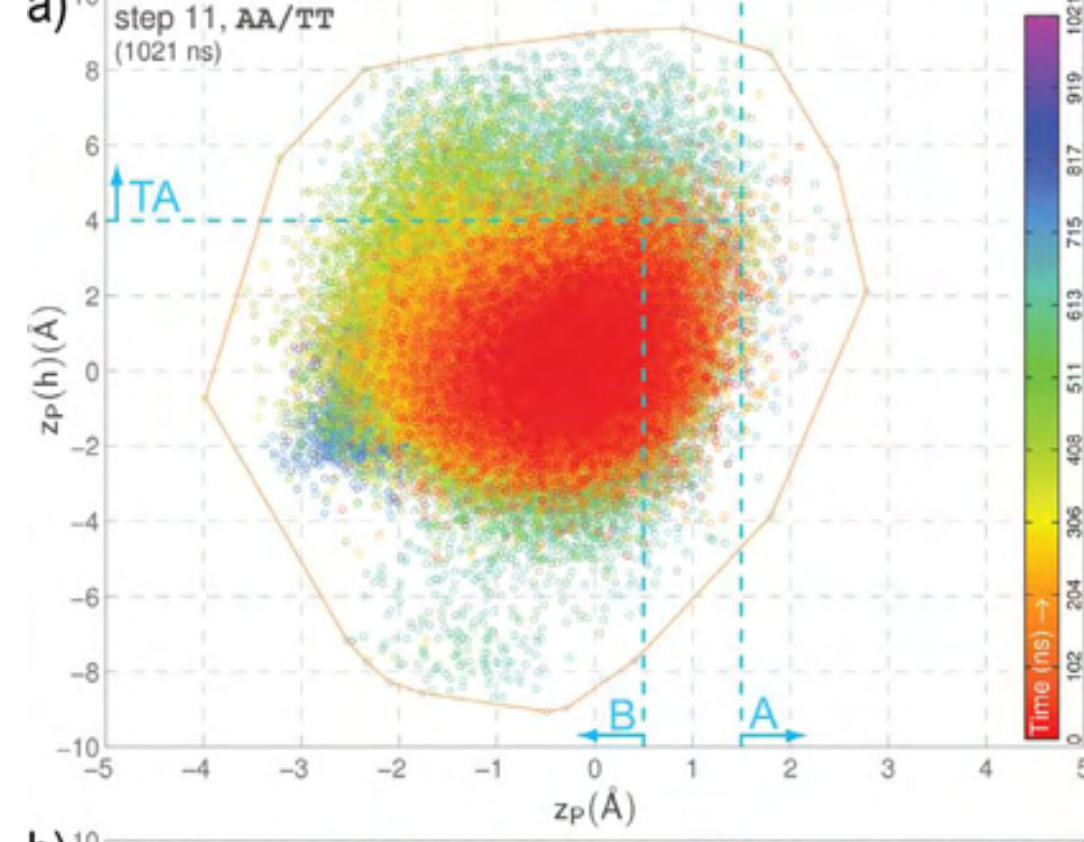

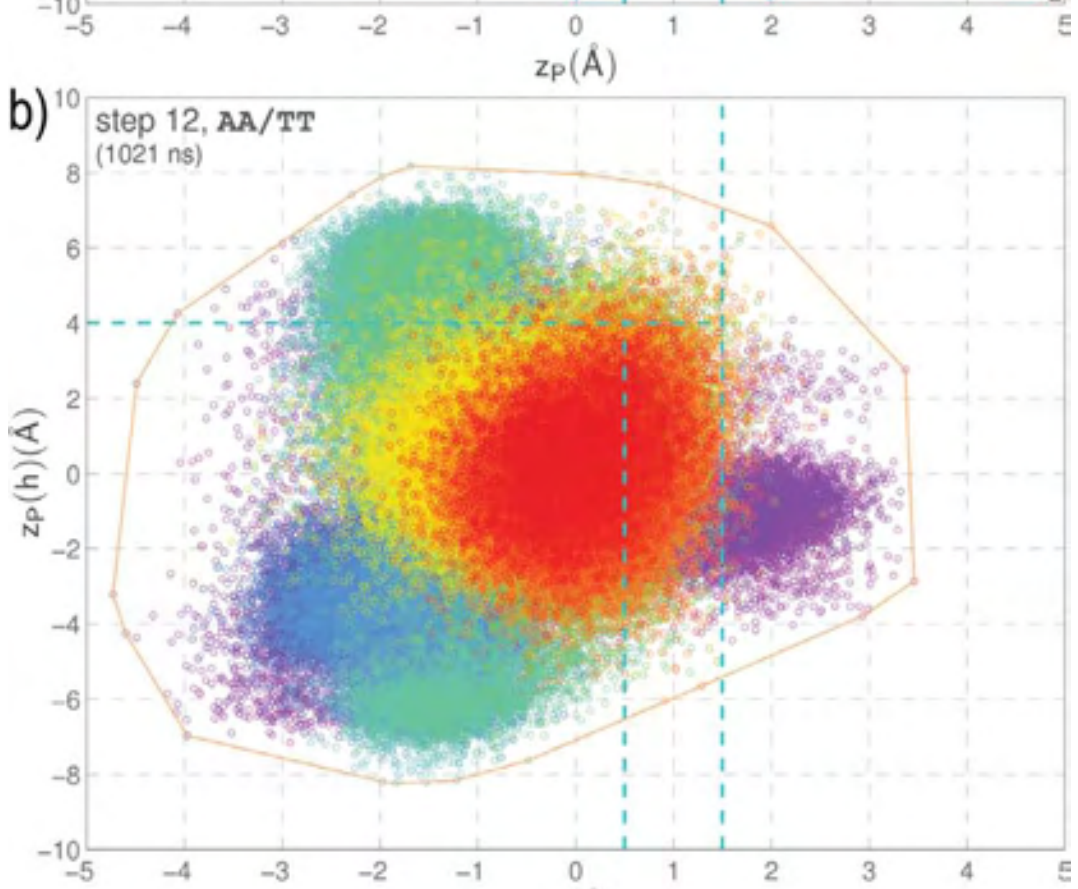

## Mura & McCammon, Figure 8

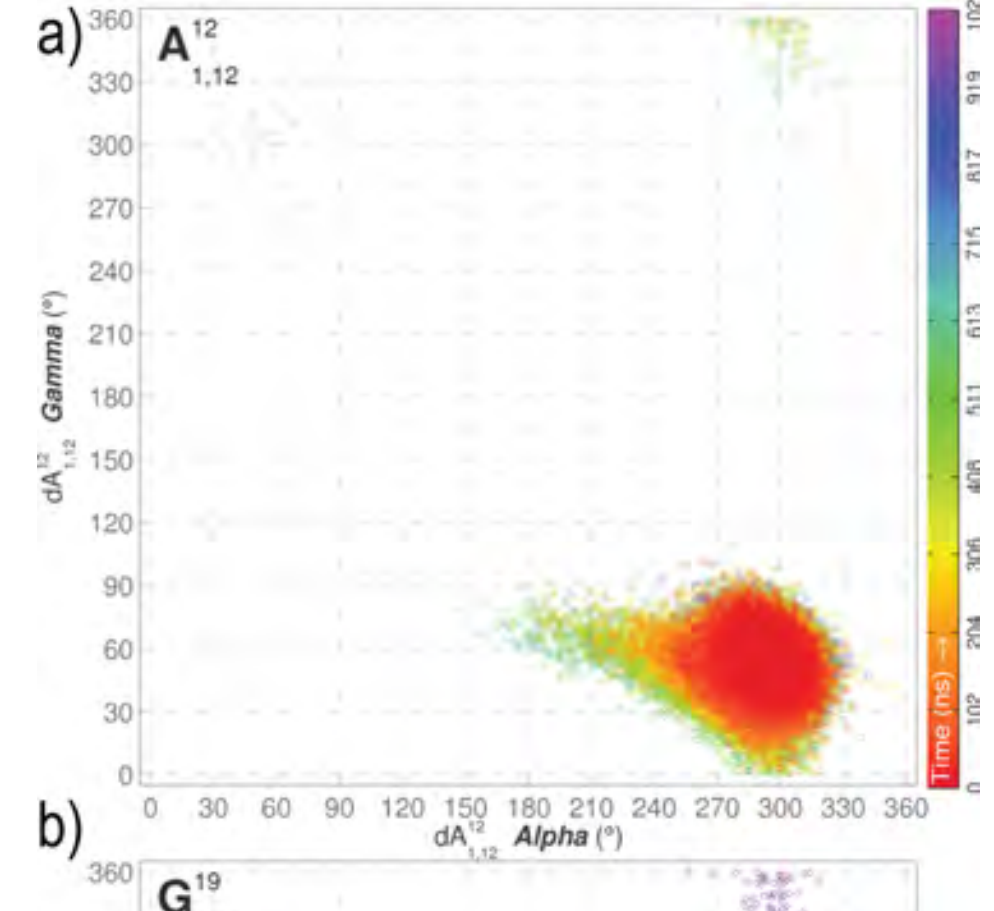

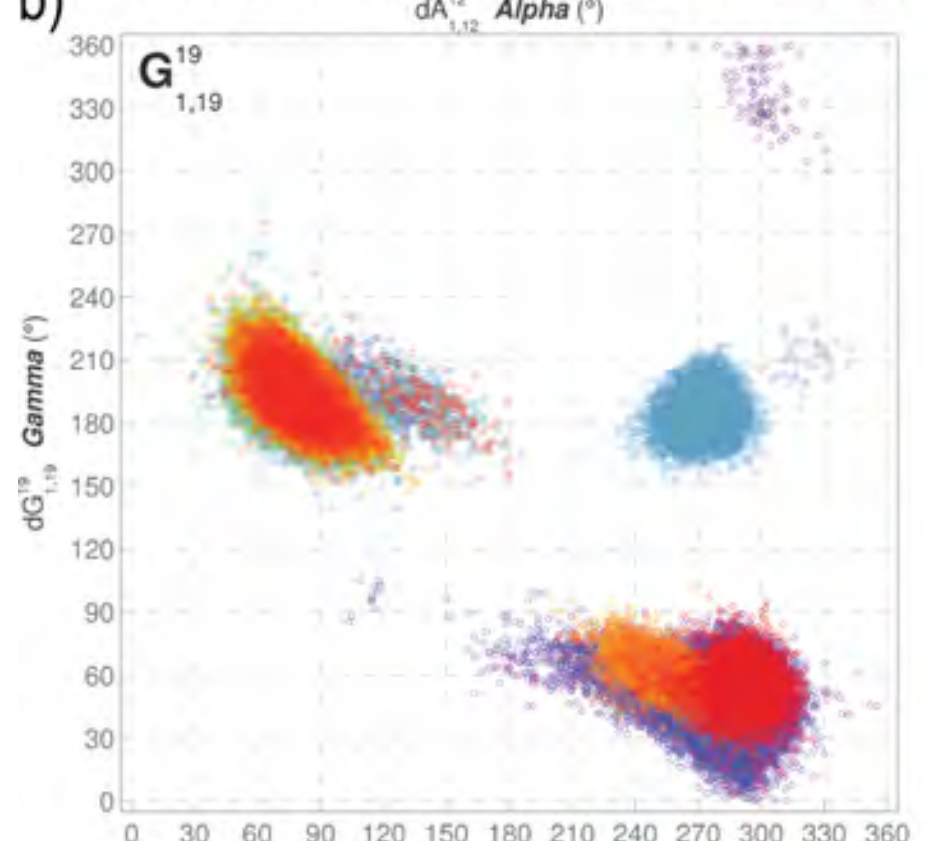

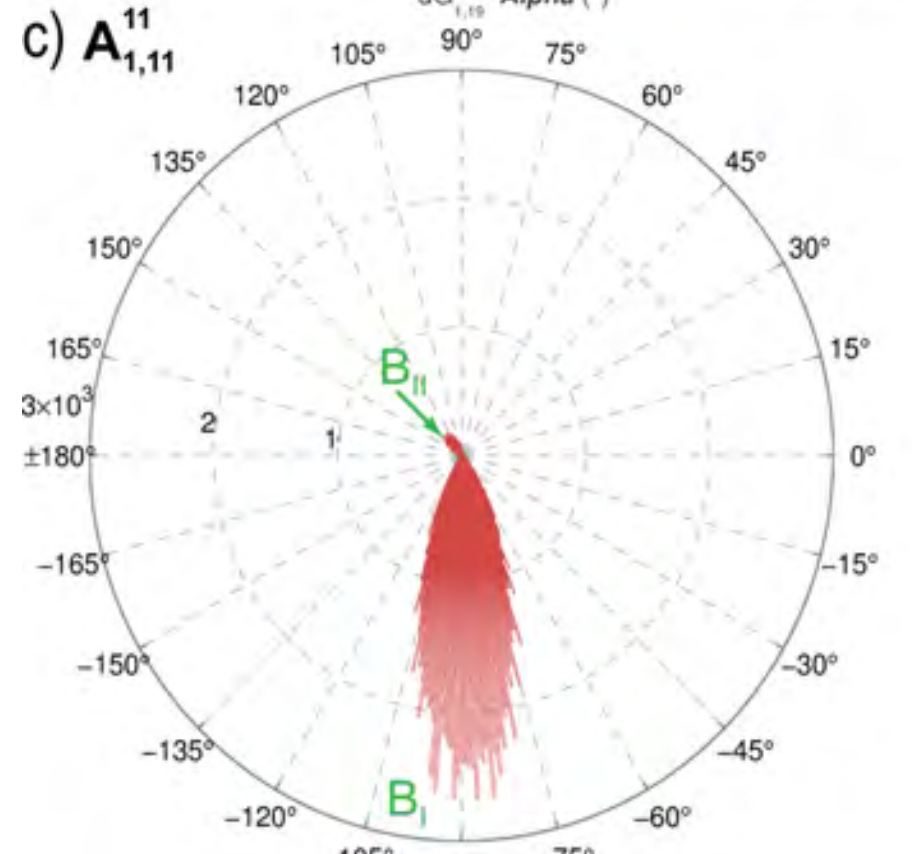